\documentclass{elsart}
\usepackage[square,comma]{natbib}
\usepackage{amsmath}
\usepackage{graphicx}
\input{tcilatex}

\begin{document}
\runauthor{W. Rolke}%
\begin{frontmatter}
\title{A Test for Equality of Distributions in High Dimensions}
\author[wolf]{Wolfgang A. Rolke}
\address[wolf]{Department of Mathematics, University of Puerto Rico - Mayag\"{u}ez, Mayag\"{u}ez, PR 00681, USA, 
\newline Postal Address: PO Box 3486, Mayag\"{u}ez, PR 00681, 
\newline Tel: (787) 255-1793, Email: wolfgang@puerto-rico.net}
\author[ang]{Angel M. L\'{o}pez}
\address[ang]{Department of Physics, University of Puerto Rico - Mayag\"{u}ez, Mayag\"{u}ez, PR 00681, USA}

\begin{abstract}
We present a method which tests whether or not two datasets (one of which could be Monte Carlo generated) might come from the same distribution. Our method works 
in arbitrarily high dimensions.
\end{abstract}
\end{frontmatter}\newpage

\section{Introduction}

Many sciences today rely heavily on the use of Monte Carlo simulation. In
High Energy Physics (HEP) for example it is used in practically every stage
of an experiment from the design of the detectors to the final analyses.
This brings up the question of the precision of the MC simulation. In other
words, how close are the probability distributions of the MC to those of the
experimental data? Testing whether two datasets come from the same
distribution is a classical problem in Statistics, and for one-dimensional
datasets a large number of methods have been developed. Tests in higher
dimensions have been proposed by Bickel \cite{Bickel} and Friedman and
Rafsky \cite{Friedman-Rafsky} . Zech \cite{Zech} discussed a test based on
the concept of minimum energy. The test proposed here belongs to a class of
consistent, asymptotically distribution free tests based on nearest
neighbors (Narsky \cite{Narsky}, Bickel and Breiman \cite{Bickel-Breiman},
Henze \cite{Henze}, and Schilling \cite{Schilling} ).

We concentrate in this paper on comparing two datasets, either both real
data or one real and one Monte Carlo, but the proposed method also allows us
to test whether a dataset comes from a known theoretical distribution as
long as we can simulate data from this distribution.

\section{The Method}

To set the stage, let's say we have observations (events) $X_{1},..,X_{n}$
from some distribution $F$, and observations $Y_{1},..,Y_{m}$ from some
distribution $G$. In the application we have in mind one of these would be
MC generated data and the other "real" data, but this is not crucial for the
following. We are interested in testing $H_{0}:F=G$ vs $H_{a}:F\neq G$. The
idea of our test is this: let's concentrate on one of the $X$ observations,
say $X_{j}$. What is its nearest neighbor, that is, the observation closest
to it? If the null hypothesis is correct and both datasets were generated by
the same probability distribution, then this nearest neighbor is equally
likely to be one of the $X$ or $Y$ observations (proportional to $n$ and $m$%
). If $F\neq G$ there should be regions in space where there are relatively
more $X$ or $Y$ observations than could be expected otherwise.

More formally, let $Z_{j}=1$ if the nearest neighbor of $X_{j}$ is from the $%
X$ data and $0$ otherwise. Then, under the null hypothesis, $Z_{j}$ is a
Bernoulli random variable with success probability $(n-1)/(n+m-1)$.
Therefore $Z=\sum_{j=1}^{n}Z_{j}$ has an approximate\ binomial distribution
with parameters $n$ and $(n-1)/(n+m-1)$. The distribution is only
approximately binomial because the $Z_{j}^{^{\prime }}$s are not
independent, but for datasets of any reasonable size the dependence is very
slight and can be ignored.

There is an immediate generalization of the test: instead of just
considering the nearest neighbor we can find the k-nearest neighbors. Now $%
Z_{j}=(Z_{j1},..,Z_{jk})$ with $Z_{ji}=1$ if the $i^{th}$ nearest neighbor
of $X_{j}$ is from the $X$ dataset,\ $0$ otherwise. Under the null
hypothesis $Z=\sum_{j=1}^{n}\sum_{i=1}^{k}Z_{ji}$ has again an approximate
binomial distribution with parameters $nk$ and $(n-1)/(n+m-1)$. (Actually, $%
\sum_{i=1}^{k}Z_{ji}$ has a hypergeometric distribution, but because we will
use a $k$ much smaller than $n$ or $m$ the difference is negligible.)

We can find the p-value of the test with%
\begin{equation*}
p=P(V\geq Z)
\end{equation*}%
where $V\sim Bin(nk,(n-1)/(n+m-1))$.

Especially if $n$ and $m$ are small or if a relatively large $k$ is desired
it would be possible to use a permutation type method to estimate the null
distribution. The idea is as follows: under $H_{0}$ all the events come from
the same distribution, so any permutation of the events will again have the
same distribution. Therefore if we join the $X$ and $Y$ events together,
shuffle them around and then split them again into $n$ and $m$\ events $%
X^{\prime }$ and $Y^{\prime }$, these are now guaranteed to have the same
distribution. Applying the k-nearest neighbor test and repeating the above
many times (say $1000$ times) will give us an estimate of the null
distribution. For more on the idea of permutation tests, see Good \cite{Good}%
. This method will achieve the correct type I error probability by
construction, but it also requires a much greater computational effort.

There are a number of choices to be made when using this method. First of
all, there is the question of which dataset should be our $X$ data. Clearly,
if $n=m$, this does not matter but it might well otherwise. Indeed, in our
application of comparing MC data to real data, we have control of the size
of the MC data although sometimes there are computational limits on its size.

Next we need to decide on $k$. Again there is no obviously optimal choice
here. Finally, we need to decide on a metric to use when finding the nearest
neighbor. If the observations differ greatly in "size" in different
dimensions, the standard Euclidean metric cannot be expected to work well
because small differences in dimensions with a large spread would overwhelm
small but significant differences in dimensions with a small spread. This
last issue we will deal with by standardizing each dimension separately,
using the mean and standard deviation of the combined $X$ and $Y$ data. If
the data come from a distribution that is severely skewed, other measures of
location and spread (such as the median and the interquartile range) could
also be used to standardize the data. In the next section we will show the
results of some mini MC studies which give some guidelines for the choices.

\section{Performance of this Method}

If we use the binomial approximation in our test, we will need to verify
that the method works, that is, that it achieves the desired type I error
probability $\alpha $. Of course, we are also interested in the power of the
test, that is, the probability to reject the null hypothesis if indeed $%
F\neq G$. In this section we will carry out several mini MC studies to
investigate these questions.

We start with the situation where there exist other methods for this test,
namely in one dimension. For comparison we will use a method that is known
to have generally very good power, the Kolmogorov-Smirnov (KS) test. In the
first simulation we generate $n=m$ observations from $1000$ to $20000$ in
steps of $1000$ each for $X$ and $Y$ from the uniform distribution on $[0,1]$%
. Because of the probability integral transform this is actually a very
general case, and similar conclusions will hold for all other continuous
distributions in one dimension. For each of these cases we use our test with 
$k=1,$ $2,$ $5$ $,10$ $,20$ $,50$ and $100$ as well as the KS test. This is
repeated $10000$ times. Figure 1 shows the results, using a nominal type I
error probability of $\alpha =5\%$.

As we see the true type I error probability is close to nominal but
increases slowly as $k$ increases. This is due to the lack of independence
between the $Z_{j}$'s. Especially if $k$ is large relative to $n$ or $m$ ,
the true type I error probability is larger than what is acceptable. Based
on this and similar simulation studies, we recommend $k=10$ if both $n$ and $%
m$ are at least $1000$, otherwise $k=0.01\min (n,m)$. Alternatively, one can
use the permutation method described above which will have the correct type
I error by construction.

Even for $k=1$ we have a slightly higher than nominal type I error
probability, about $5.5\%$ if $\alpha =5\%$. This is partly due to the
dependence between the $Z_{j}$'s, and partly to the discrete nature of the
binomial distribution. For example, if we defined the p-value as $P(V\geq
Z-1)$ we would get a true type I error probability slightly smaller than the
nominal one. In any case, we believe this difference to be acceptable.

Next a simulation to study the power of the test, again in one dimension. We
use the uniform on $[0,1]$ distribution for $F$ and the uniform on $%
[0,\theta ]$ for $G$, where $\theta $ goes from $1$ to $1.1$ We generate $%
n=m=1000$ events and apply our test with $k=1$, $5$, $15$ and $25$ as well
as the KS test. This is repeated $10000$ times. The result is shown in
Figure 2. Clearly, the higher the $k$ the better the power of the test. In
fact, for this example already with $k=15$ the test has better power than
the KS test! Generally speaking, in one dimension, our test has power
somewhat inferior to the KS test. However, our test is not meant to be used
in one dimension. It is encouraging to find that it does fairly well even in
that situation.

How should one choose the size of the Monte Carlo dataset relative to the
size of the true data? In the next simulation we generate $n=1000$ events
from a uniform $[0,1]$ and assume this to be the real data. Then we generate 
$m$ events from the same distribution, with $m$ going from $50$ to $2500$.
In Figure 3 we show the results which indicate that the MC dataset should
have the same size as the real data because in that case the true type I
error probability is about the same as the nominal one. This agrees with
general statistical experience which suggests that, in two-sample problems,
equal sample sizes are often preferable.

Finally we present a multi-dimensional example. We generate $n=m=1000$
events. The $F$ distribution is a multivariate standard normal in $9$
dimensions, and the $G$ distribution a multivariate normal in $9$ dimensions
with means $0$, standard deviations $1$ and correlation coefficients $%
cor(X_{i},X_{j})=\rho $ if $|i-j|=1$ and $0$ if $|i-j|>1$, where $\rho $
goes from $0$ to $0.5$. This example illustrates the need for a test in
higher dimensions. Here all the marginals are standard normals, and any
one-dimension-at-a-time method is certain to miss the difference between $F$
and $G$. This is shown in Figure 4 where with $k=10$ we reject the null
hypothesis quite easily (at $\rho =0.5$) whereas the KS test applied at any
of the marginals fails completely.

\section{Implementation}

A C++ routine that carries out the test is available from one of the authors
at http://math.uprm.edu/\symbol{126}wrolke/. It allows the use of the
binomial approximation as well as the permutation method. It uses a simple
search for the k-nearest neighbors. k-NN searching is a well known problem
in computer science, and more sophisticated and faster routines exist and
could also be used in combination with our code. (See, for example,
Friedman, Baskett and Shustek \cite{Friedman et al} .)

\section{Summary}

We describe a test for the equality of distributions of two datasets in
higher dimensions. The test is conceptually simple and does not suffer from
the curse of dimensionality. Simulation studies show that it approximately
achieves the desired type I error probability, or does so exactly at a
higher computational cost. They also show that this test is capable to
detect differences between the distributions only "visible" in higher
dimensions.

\section{Appendix}

\begin{figure}
\centering  
\includegraphics[width=0.90\textwidth]{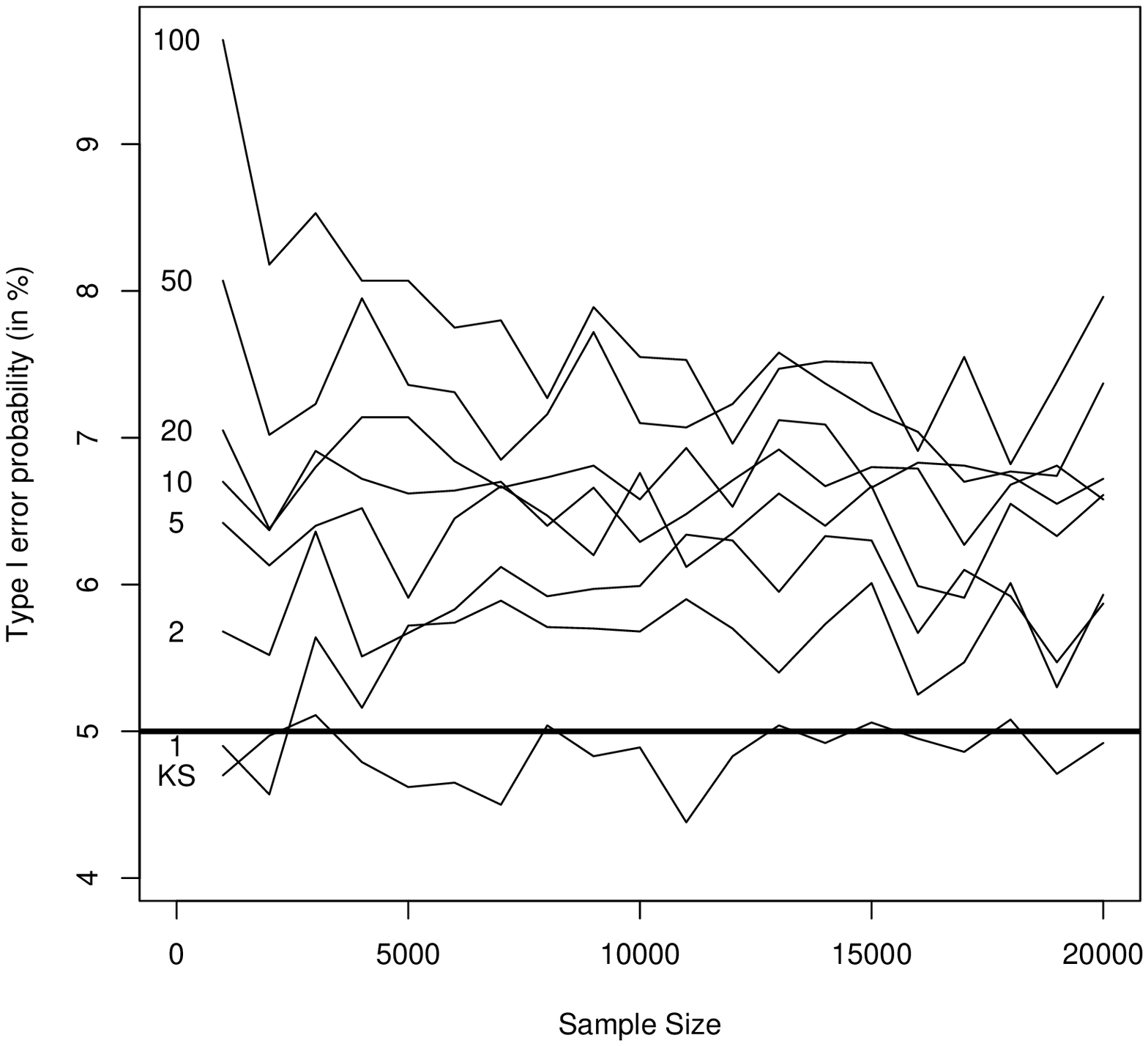}  
\caption{The true type I error
probability in a one dimensional example as a function of sample size. $X$
and $Y$ have a uniform distribution on $[0,1]$. $n=m$, and the nominal type
I error probability is $5\%$. Each curve represents a different $k$ value.
The KS test gives a flat line at $5\%$.
}
\label{fig:fig1}
\end{figure}

\begin{figure}
   \centering  
          \includegraphics{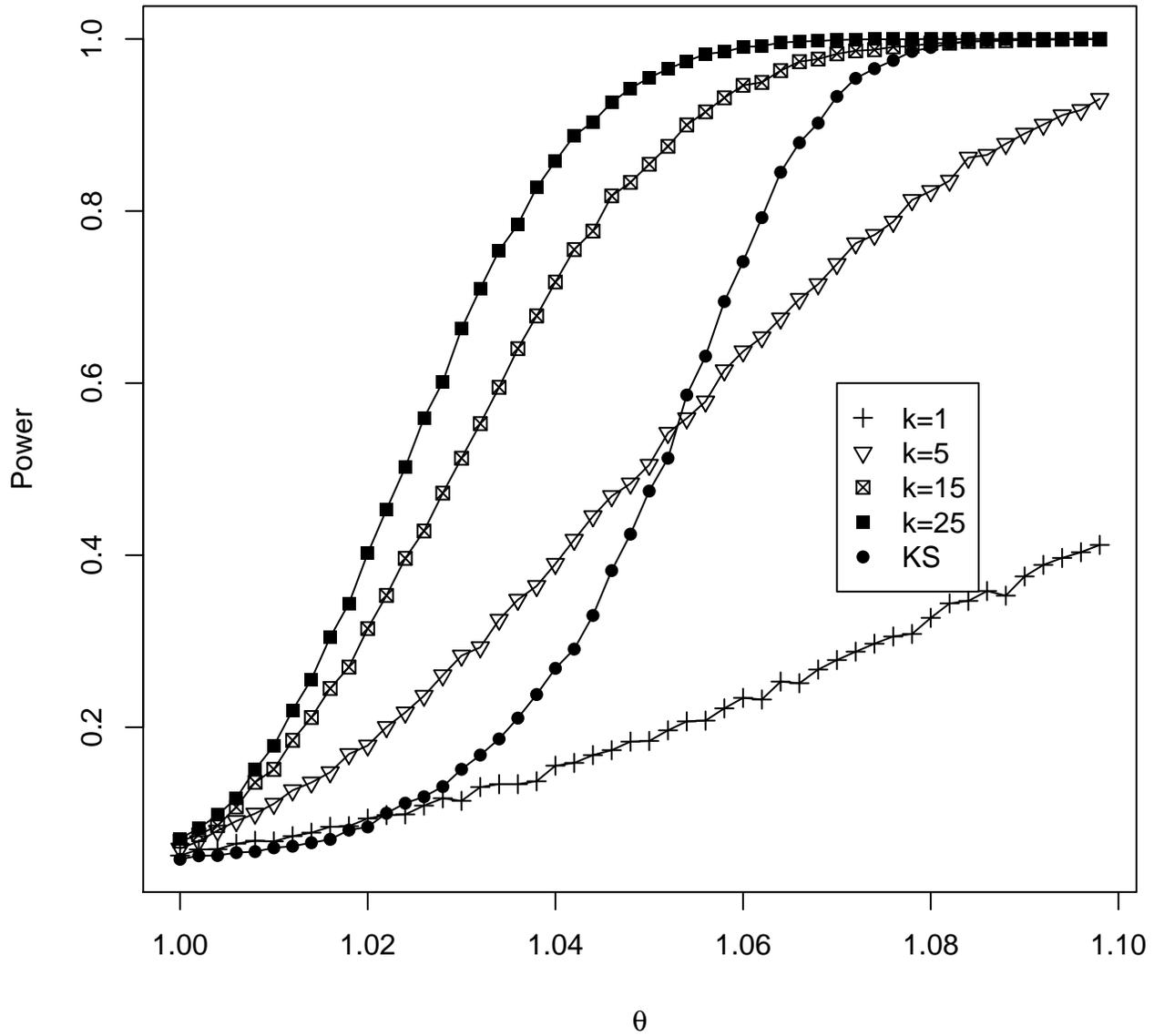}  
           \caption{$F=U[0,1]$,  $G=U[0,\theta ]$ where $\theta $ goes from $1$ to $1.1$ We generate 
$n=m=1000$ events. $k=1$, $5$, $15$ and $25$.}
\label{fig:fig2}
\end{figure}

\begin{figure}[tbp]
\centering  \includegraphics[width=0.90\textwidth]{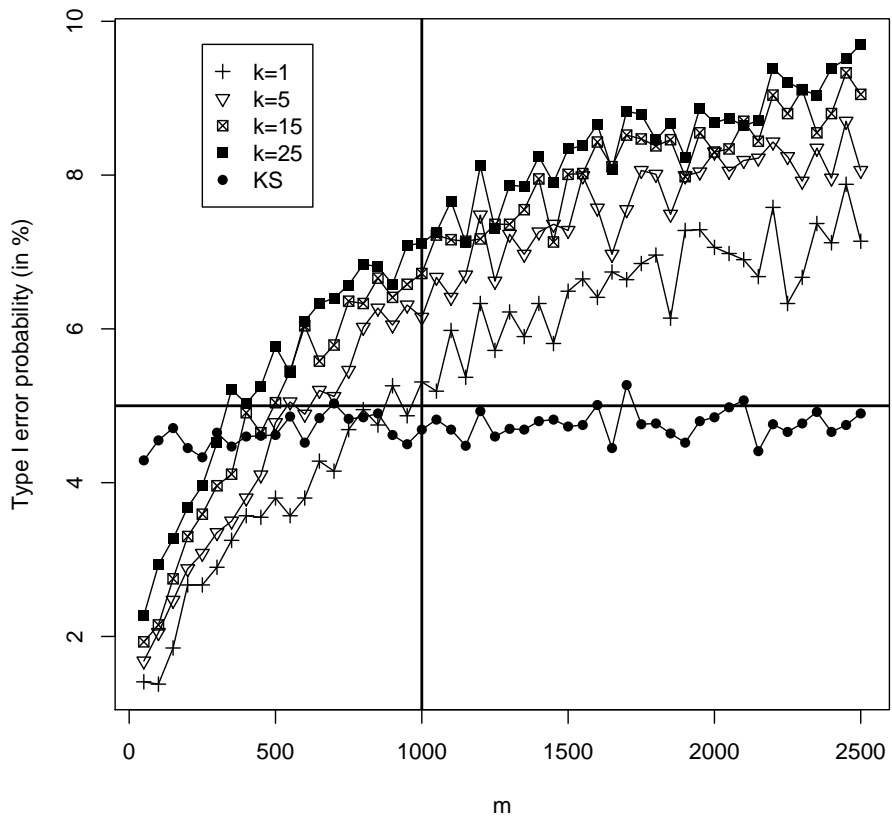}  
\caption{$n=1000$ from $%
F=U[0,1]$, $m$ goes from $50$ to $2500$, $G=F$.
}
\label{fig:fig3}
\end{figure}

\begin{figure}[tbp]
\centering  \includegraphics[width=0.90\textwidth]{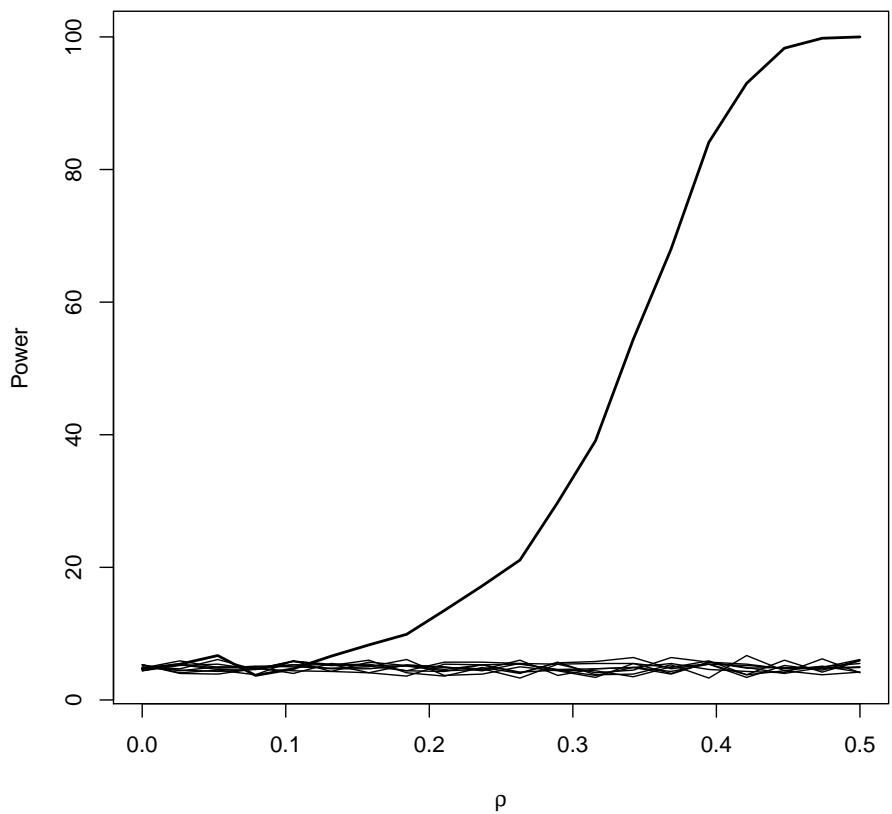}  
\caption{$n=m=1000$. The $F$
distribution is a multivariate standard normal in $9$ dimensions, and the $G$
distribution a multivariate normal in $9$ dimensions with means $0$,
standard deviations $1$ and correlation coefficients $cor(X_{i},X_{j})=%
\protect\rho $ if $|i-j|=1$ and $0$ if $|i-j|>1$, where $\protect\rho $ goes
from $0$ to $0.5$. The multidimensional test has much better power than any
of the one-dimensional KS tests.
}
\label{fig:fig4}
\end{figure}

\end{document}